\documentclass{article}
\usepackage{graphicx} 
\usepackage[utf8]{inputenc}
\usepackage{authblk}
\usepackage{caption}
\usepackage{hyperref}
\usepackage{mathtools, nccmath}
\usepackage{amssymb}
\usepackage{geometry}
\title{Use of spatiotemporal couplings and an axiparabola to control axial energy deposition velocity}
\author[1,*]{Aaron Liberman}
\author[2]{Ronan Lahaye}
\author[1,2]{Slava Smartsev}
\author[1]{Sheroy Tata}
\author[1]{Salome Benracassa}
\author[1]{Eitan Levine}
\author[2]{Cedric Thaury}
\author[1]{Victor Malka}

\affil[1]{Department of Physics of Complex Systems, Weizmann Institute of Science, Rehovot 7610001, Israel}
\affil[2]{LOA, CNRS, Ecole Polytechnique, ENSTA Paris, Institut Polytechnique de Paris,  Palaiseau, France}

\affil[*]{Corresponding author: aaronrafael.liberman@weizmann.ac.il}
\date{}

\usepackage{lineno}
\usepackage{float}

\begin{document}

\maketitle





\renewcommand{\topfraction}{.85}
\renewcommand{\bottomfraction}{.7}
\renewcommand{\textfraction}{.15}
\renewcommand{\floatpagefraction}{.66}
\renewcommand{\dbltopfraction}{.66}
\renewcommand{\dblfloatpagefraction}{.66}
\setcounter{topnumber}{9}
\setcounter{bottomnumber}{9}
\setcounter{totalnumber}{20}
\setcounter{dbltopnumber}{9}

\begin{abstract}

This paper presents the first experimental realization of a scheme that allows for the tuning of the axial energy deposition velocity of a focal spot with relativistic intensity. By combining a tunable pulse-front curvature with the axial energy deposition characteristics of the axiparabola, an aspheric optical element, this system allows for controlling the dynamics of laser-wakefield accelerators. We demonstrate the ability to modify the axial energy deposition velocity of 100 TW laser pulses to be superluminal or subluminal. The experimental results are supported by theoretical calculations and simulations, strengthening the case for the axiparabola as a pertinent strategy to achieve dephasingless acceleration.

\end{abstract}

\maketitle

\section{Introduction}

With the proliferation of femtosecond, multi-terawatt pulses, high power laser systems have become an essential tool in scientific research. In particular, such laser systems, when coupled to plasma, can produce a compact particle accelerator, capable of generating high quality proton, electron, and x-ray beams \cite{Mangles_Nature_2004,Geddes_Nature_2004,Faure_Nature_2004,Malka_NaturePhysics_2008}. These laser-wakefield accelerators (LWFAs) \cite{Tajima_PRL_1979} can produce GeV energy scale electrons in centimeter-sized acceleration lengths \cite{Gonsalves_PRL_2019}. LWFAs promise to contribute to cancer therapy treatments \cite{Raschke_ScientificReports_2016}, radiobiology \cite{Cavallone_PhysicaMedica_2019}, and material science \cite{Glinec_PRL_2005} and as well as to next-generation light sources, such as FELs \cite{Wang_Nature_2021}. There remain, however, a number of limitations that prevent the achievement of ever higher energies \cite{Esarey_ReviewOfModernPhysics_2009}. These include the tendency of the laser to diffract \cite{Esarey_ReviewOfModernPhysics_2009} -  requiring guiding mechanisms to remain in focus \cite{Leemans_PRL_2014} - and the dephasing of the electrons from the wakefield \cite{Esarey_ReviewOfModernPhysics_2009}, potentially ending the acceleration before the laser is depleted \cite{Xie_PhysicsOfPlasmas_2007}.

The axiparabola, a long-focal-depth reflective optical element that produces a quasi-Bessel beam \cite{Smartsev_OpticsLetters_2019}, has generated interest in its potential to both overcome beam diffraction and electron dephasing \cite{Palastro_PRL_2020,Caizergues_NaturePhotonics_2020}. The former is accomplished by the diffraction-free propagation properties of the Bessel beam, which have been used to generate a high-quality waveguide for LWFA \cite{Oubrerie_Light_2022}. The latter, meanwhile, is handled through a combination of the dynamics imposed by the axiparabola itself and a manipulation of the pulse-front curvature (PFC) of the incoming beam \cite{Palastro_PRL_2020,Caizergues_NaturePhotonics_2020}. Through this combination, the axial energy deposition velocity can be tuned from subluminal to superluminal, allowing the wakefield to be phase-locked to the electron beam. This paper presents the first experimental realization of a system that promises to take advantage of the dephasingless acceleration properties of the axiparabola. The manipulation of the energy deposition velocity is acheived by combining an axiparabola and a refractive doublet that modifies the PFC of the beam. The experiment was performed with the HIGGINS 100 TW laser system at the Weizmann Institute of Science, which produces nearly top-hat pulses of 30 fs with a 50 mm diameter   \cite{Kroupp_MRE_2022}.

While superluminal energy deposition has been demonstrated in the past \cite{Mugnai_PRL_2000,Alexeev_PRL_2002,Froula_NaturePhotonics_2018,Jolly_OpticsExpress_2020}, this is the first experiment which shows its feasibility using the axiparabola and, thus, presents a roadmap for further optimization towards the eventual goal of dephasingless laser-wakefield acceleration. Furthermore, the axiparabola approach does not require adding significant chirp to the pulse, such as the flying focus technique \cite{Sainte-Marie_Optica_2017,Froula_NaturePhotonics_2018,Jolly_OpticsExpress_2020}. The axiparabola-based method is a promising, experimentally simpler way to achieve dephasingless acceleration.

\section{Methods}

\subsection*{Axiparabola} 
The axiparabola is a long-focal-depth optical element that causes a radially dependent focal length, $f(r)$, with different annular segments focusing at different points on the focal line over a segment known as the focal depth. A detailed description of the axiparabola can be found in \cite{Smartsev_OpticsLetters_2019,Oubrerie_JoO_2022}. 

By causing light at different radial distances to have different path lengths to the optical axis, the axiparabola introduces a radially dependent focusing time, $t(r)$, and thus modifies the velocity of the on-axis intensity deposition, $v_z = df/dt$ \cite{Caizergues_NaturePhotonics_2020, Palastro_PRL_2020}. Here, $z = f-f_0$ where $f_0$ is the nominal focal length of the axiparabola. Using ray-optics and paraxial approximations the velocity is  $v_z/c = 1 + r^2/2f^2$ \cite{Caizergues_NaturePhotonics_2020}. 
If the pulse incident on the axiparabola has a radial delay $\tau(r)$ the velocity is modified \cite{Oubrerie_JoO_2022}:
\begin{equation}
    \frac{v_z^{'}}{c} =\frac{v_z}{c}\bigg(1-\frac{v_z}{c} c \frac{d\tau}{dr}\frac{dr}{dz}\bigg)
\end{equation}
This experiment used an axiparabola with the focal line distribution: $ f(r) = f_0 + \delta (r/R)^2$, where $\delta>0$ is the focal depth and $R$ is the total radius of the pulse. For such an axiparabola, and with a radial delay caused by pulse-front curvature of $\tau(r) = \alpha r^2$, the modified velocity, up to second order terms becomes:

\begin{equation}
    \frac{v_z^{'}}{c} = 1 - \frac{c\alpha R^2}{\delta}+\bigg(\frac{R^2}{2\delta f_0} - \frac{c\alpha R^4}{\delta^2 f_0^2}\bigg)z-\bigg(\bigg[\frac{R^2}{2\delta f_0^2} - \frac{c\alpha R^4}{\delta^2 f_0^2}\bigg]\frac{2}{f_0} -  \frac{c\alpha R^6}{2\delta^3 f_0^4}\bigg)z^2 \label{eq:full_expression}
\end{equation}

\subsection*{Pulse Front Curvature Measurement and Control}

\noindent The pulse-front curvature of the HIGGINS laser was measured using far-field beamlet cross-correlation (FFBCC) \cite{Smartsev_JoO_2022}, a self-referenced technique which utilizes far-field interferometry and inverse Fourier transform spectroscopy. The method works by scanning the delay between two beamlets and using the interference obtained by focusing them using a parabola, in the case of this PFC measurement a 2 m off-axis parabola, to extract the relative pulse delay. Repeating this for different sections of the beam maps the pulse-front delay \cite{Smartsev_JoO_2022}. 


The PFC control doublet, a specially designed refractive doublet, was placed inside the final telescope expansion stage of the laser. The doublet was designed to cancel out the PFC which is introduced to the laser system due to the refractive telescopes used to expand the beam \cite{Netz_AppliedPhysicsB_2000}. Details of the doublet are provided in \cite{Smartsev_JoO_2022}. Since the doublet's effect scales with the size of the beam impinging on the doublet \cite{Kabacinski_2021}, by changing the doublet's position inside of the telescope, the PFC of the beam can be controlled in amplitude and direction without changing the focal plane and with minimal added aberrations.  

\begin{figure}[t]
		\centering
		\captionsetup{width=0.8\linewidth}
		\includegraphics[width=0.6\linewidth]{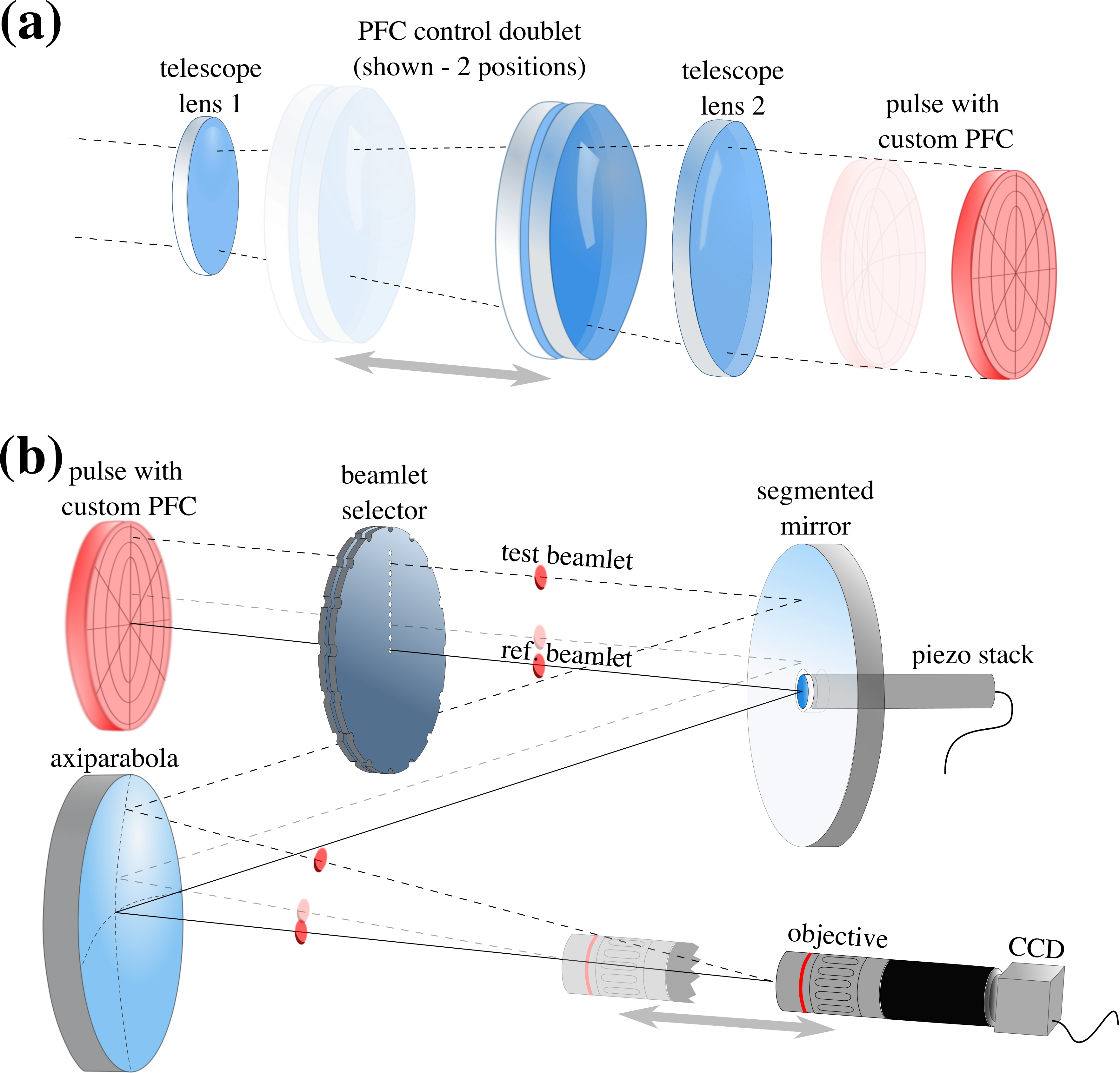}
		\caption{\label{fig:Setup} \small (a) Setup of PFC control doublet inside the final beam expansion stage. Shown in two positions and with two resultant PFC values for the beam. (b) Group velocity measurement setup. Beam impinges on selector, reference and test beamlets pass through. Segmented mirror gives tunable delay to the reference beamlet. Both beamlets are focused by axiparabola and the interference is imaged on CCD through an objective. Shown for two radial values of the test beamlet. }
\end{figure}

\subsection*{PFC Ray-Optics Simulation}

\noindent The PFC of the laser chain was simulated using the Zemax OpticStudio ray-tracing software. It takes into account the final three beam expansion stages of the Weizmann laser, which contribute the most to the PFC. A fourth, perfect dispersionless telescope, is used to approximate the propagation in free space. A custom macro \cite{Slava_Thesis} was used to calculate the pulse-front delay, which uses ray tracing of frequencies around the central wavelength (800 nm) for each position of the beam. By calculating the spatially dependent spectral phase, the value of the PFC can be extracted. Another macro \cite{Slava_Thesis}, based on fast Fourier transforms, simulated the intensity of the axiparabola focused beam over the focal depth. 

\subsection*{Velocity Measurement}

\noindent The group velocity measurement, an overview of which is shown in figure \ref{fig:Setup}, was performed in air and was based on a modified version of FFBCC. The deposition velocity measurement used a specially designed beamlet selector mask that selects a test beamlet from a radially and angularly tunable section of the beam and selects the central, reference beamlet. A variable delay is added to the reference beamlet by impinging the beamlets onto a segmented mirror whose central section is attached to a piezo actuator. The beamlets were then focused by the axiparabola, and the resultant interference was imaged using an objective and CCD. The imaging system was moved to coincide with the focus of the test and reference beamlets, and a delay scan was performed for four radial offset values. The resulting value for the delay difference for each radial slice was extracted by finding the relative delay needed to achieve maximum interference of the two beamlets. Since the reference beamlet at $r=0$ corresponds to the place where the average of the $\Vec{k}$s for all other points is 0, the relative delay of the test beamlet is taken as deviation from a luminal arrival time. The data for the different test beamlet values was fitted and the derivative of this fit function yields the difference between the axial propagation velocity and the speed of light. Measurements were also done to verify that the objective introduced negligible chromatic dispersion, and thus did not introduce radial delay.

Experimental data was first taken without the PFC control doublet in the laser chain. The control doublet was then introduced. By manipulating the PFC of the beam incoming on the axiparabola, the doublet changes the relative timing of arrival of different radial segments of the laser to the axiparabola, and thus the timing of the axial energy deposition.

\section{RESULTS}
\subsection*{Axiparabola Focal Spot}
\label{subsec:Axi_res}

\begin{figure}[b]
		\centering
		\captionsetup{width=\linewidth}
		\includegraphics[width=\linewidth]{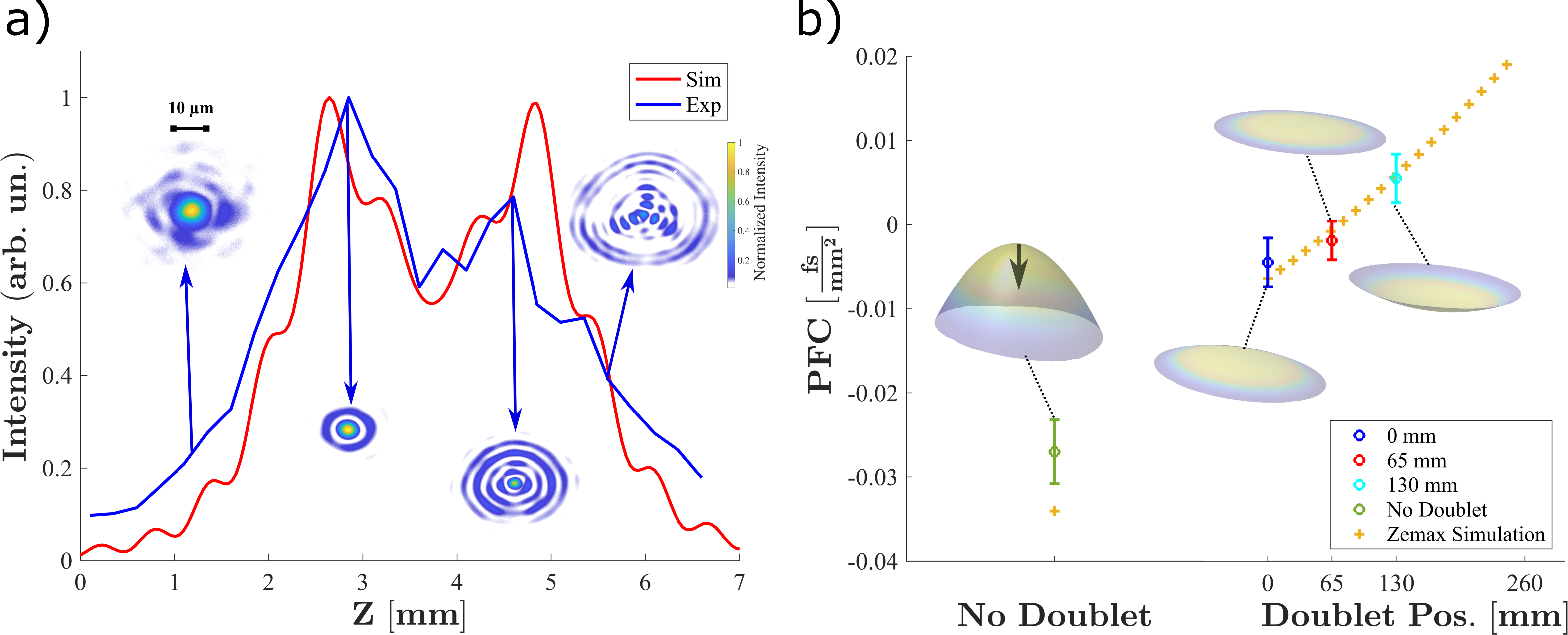}
		\caption{\label{fig:Combined}\small (a) Intensity over focal depth. Zemax simulation (red) and experimental data (blue) shown. Inset images show 2D focal spot data at specified points. (b) Zemax simulated PFC (orange) and experimentally measured PFC for 0 mm (blue), 65 mm (red) and 130 mm (cyan) doublet positions and the no doublet case (green). The insets show the measured pulse-front delays. Black arrow shows the propagation direction of pulse.}
\end{figure}

\noindent The axiparabola used was a 480 mm nominal focal length, 5 mm focal depth axiparabola designed to have a quasi-constant intensity over the focal depth. Due to interference effects, the intensity is actually a double peak along the focal depth, best approximating the constant intensity. Figure \ref{fig:Combined} (a) shows both the Zemax OpticStudio simulation (red) and the experimentally obtained (blue) intensity of the axiparabola central spot over the focal depth. Insets show the measured focal spot at different z values, illustrating the development of the Bessel ring structures. The second peak of the intensity profile is experimentally reduced relative to the simulation due to spatial aberrations. These appear because the end of the focal depth is generated by large aperture annuli,  and is thus more susceptible to aberrations \cite{Smartsev_OpticsLetters_2019}.

\subsection*{PFC Manipulation}

\noindent In the experiment, four doublet positions were used: 0 mm, 65 mm, 130 mm, and 260 mm. They span the range of the doublet from the beginning to the end of the telescope it was placed in. The modification of the PFC by the doublet increases with larger values of the position inside the telescope as the beam's size grows. Data without the doublet was also taken.  Figure \ref{fig:Combined} (b) shows the Zemax simulated PFC (orange) as well as the experimentally obtained values for the 0 mm (blue), 65 mm (red) and 130 mm (cyan) doublet positions and the no doublet case (green). The insets show the measured pulse-front delays. As shown in the figure, the range of PFC allowed by moving the doublet in the telescope gives the ability to partially and fully suppress, and even invert, the PFC inherent in the refractive beam expansion. The simulations for the doublet-in cases agree with the experimentally obtained values, while the no-doublet case deviates slightly, likely due to experimental errors.

\subsection*{Velocity Measurement}

\begin{figure}[h]
		\centering
		\captionsetup{width=\linewidth}
		\includegraphics[width=\linewidth]{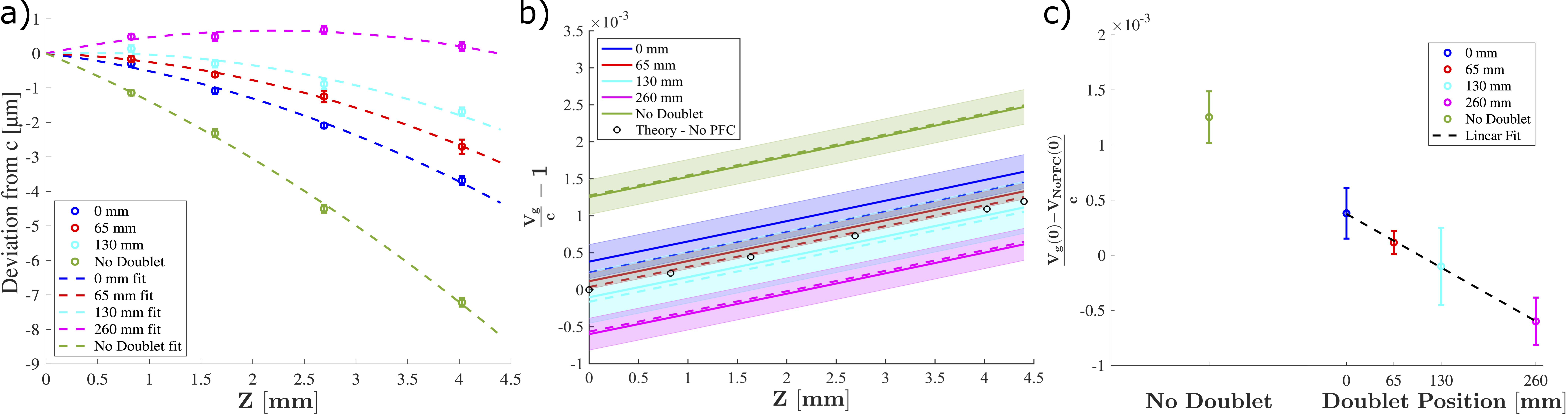}
		\caption{\label{fig:Axi_Velocity} \small (a) Deviation of annular sections of the axiparabola from luminal propagation for the no doublet (green), doublet 0 mm (blue), 65 mm (red), 130 mm (cyan), and 260 mm (magenta) cases. Dotted line shows parabolic fit. (b) Negative derivative of the fit functions shown in part a. Shaded sections show error bars, dotted lines show theory fits, and black circles show no-PFC theory. (c) Initial deviation from no PFC velocity with different doublet positions. Black dotted line is a linear fit.   }
\end{figure}

\noindent The results of the scan of the relative delay between luminal propagation and the arrival of the annular sections of the axiparabola are shown in figure \ref{fig:Axi_Velocity} (a). The results are shown for data taken at focal positions $z = 0.8, 1.6, 2.7, 4.0\ \text{mm}$, which correspond to the focusing points of the axiparabola annular sections at radial values $r =10.3,14.5,18.6,22.8\ \text{mm}$, respectively. These are the radial values of the test beamlets selected by the beamlet selector mask. Data is shown for the case where no PFC control doublet is in the laser chain (green), as well as when the doublet is at the 0 mm (blue), 65 mm (red), 130 mm (cyan), and 260 mm (magenta) positions inside of the final beam expander. 

For each case, the deviation data was fitted using a third order polynomial fit function:
\begin{equation}
    \text{d}(z) = p_1z^3 +p_2z^2+p_3z+p_4 \label{eq:fit}
\end{equation} where d is the deviation, to match the integral of the velocity equation shown in equation \ref{eq:full_expression}. Since each experimental fit had few points and the expectation is that the dominant contribution to the velocity change will come from the $p_3z$ term, the values of $p_1$ and $p_2$ were taken to be the theoretical values given by the Zemax simulated PFC shown in figure \ref{fig:Combined} (b) and the fit was used to obtain the values of $p_3$ and $p_4$. The experimental error bars shown stem from the error inherent in the FFBCC measurement.

From the derivative of the fit function in equation \ref{eq:fit}, the deviation of group velocity from luminal velocity can be obtained. 
The results are shown in figure \ref{fig:Axi_Velocity} (b). 
As expected, for the no PFC control doublet case (green), the PFC inherent in the refractive beam expansion causes the most superluminal propagation velocity. When the doublet is introduced at 0 mm (blue), the starting position of the velocity is reduced. As the doublet is moved forward in the telescope to 65 mm (red), 130 mm (cyan), and 260 mm (magenta), the starting value of the velocity is further reduced. As the PFC of the beam is flipped in direction, between the doublet 65 mm and double 130 mm positions, the starting velocity switched from superluminal to subluminal, as is expected from equation \ref{eq:full_expression}. The shaded regions are the error bars for each measurement. The dotted line of each color, represents the theoretical expectation of the velocity, obtained from equation \ref{eq:full_expression}, using the Zemax values of the PFC for each doublet position. The theory agrees quite well with the experimental data, falling well within the error boundaries in each case. The figure also shows the theoretical value for no PFC, drawn in black circles, which, as expected, falls between the doublet 65 mm and doublet 130 mm positions, which sit on either side of the PFC direction change. 

Figure \ref{fig:Axi_Velocity}(c) shows the difference in the starting ($z=0$) velocity of the different doublet positions from the velocity with no PFC. The left side shows the no doublet case (green) while the right shows a graph with the different doublet positions (same coloring as before). The black dotted line shows the linear fit of the experimentally measured data as a function of doublet position. This figure highlights the dynamic range in velocity control afforded by the PFC manipulation.  

\section{Conclusion}
This paper presents the first experimental realization of an axiparabola based scheme for tuning the axial deposition velocity. The dynamic range shown in the velocity measurements allows for the easy transition between  superluminal, luminal, and subluminal velocities. These results confirm that the axiparabola behaves as expected, both in focal spot intensity over the focal depth and in velocity. With the axiparabola characterized and the velocity manipulation demonstrated, this setup is now ready to facilitate the first high-power axiparabola based attempts at dephasingless laser-wakefield acceleration.

\bibliographystyle{ieeetr} 
\bibliography{paper} 

\end{document}